\documentclass[twocolumn,showpacs,preprintnumbers,amsmath,amssymb,prb,superscriptaddress]{revtex4}

\usepackage{epstopdf}
\usepackage{graphicx}
\usepackage{multirow}

\def\BSIO{$\rm Sr_3Ir_2O_7$}
\def\SIO{$\rm Sr_2IrO_4$}
\def\NIO{$\rm Na_2IrO_3$}
\def\LIO{$\rm Li_2IrO_3$}

\def\jeff{$j_{\rm{eff}}$}

\begin{document}

\title{Magnetic excitation spectrum of Na$_2$IrO$_3$ probed with resonant inelastic x-ray scattering}

\author{H.~Gretarsson}
\author{J. P. Clancy}
\affiliation{Department of Physics, University of Toronto, 60
St.~George St., Toronto, Ontario, M5S 1A7, Canada}
\author{Yogesh Singh}
\affiliation{Indian Institute of Science Education and Research Mohali, Sector 81, SAS
Nagar, Manauli PO 140306, India }
\author{P. Gegenwart}
\affiliation{Physikalisches Institut, Georg-August-Universit\"at
G\"ottingen, D-37077, G\"ottingen, Germany}
\author{J.~P. Hill}
\affiliation{CMP\&MS Department, Brookhaven National Laboratory,
Upton, New York 11973, USA}
\author{Jungho~Kim}
\author{M.~H. Upton}
\author{A. H. Said}
\author{D.~Casa}
\author{T.~Gog}
\affiliation{Advanced Photon Source, Argonne National Laboratory,
Argonne, Illinois 60439, USA}
\author{Young-June Kim}
\email{yjkim@physics.utoronto.ca} \affiliation{Department of
Physics, University of Toronto, 60 St.~George St., Toronto, Ontario,
M5S 1A7, Canada}

\date{\today}

\begin{abstract}
The low energy excitations in \NIO\ have been investigated using resonant inelastic x-ray scattering (RIXS). A magnetic excitation branch can be resolved, whose dispersion reaches a maximum energy of about 35 meV at the $\Gamma$-point. The momentum dependence of the excitation energy is much larger along the $\Gamma-X$ direction compared to that along the $\Gamma-Y$ direction. The observed dispersion relation is consistent with a recent theoretical prediction based on Heisenberg-Kitaev model. At high temperatures, we find large contributions from lattice vibrational modes to our RIXS spectra, suggesting that a strong electron-lattice coupling is present in \NIO.
 \end{abstract}

\maketitle

%%%%%%%%%%%%%%%%%%
%% Introduction %%
%%%%%%%%%%%%%%%%%%

The physics of iridates has drawn considerable attention recently. \cite{Okamoto2007, BJKIM2008, BJKim2009, Jackeli2009, Shitade2009, Pesin2010, Jiri2010,Khaliullin2012,Sr2IrO4-RIXS,Sr3Ir2O7-RIXS, Clancy2012,Subhro} One of the reasons for this surge of interest is the fact that the local magnetic moment arises from a spin-orbit coupled \jeff=1/2 state rather than a spin only state with quenched orbital moment as usually found in 3d transition metal compounds. One of the consequences of the \jeff=1/2 ground state is that the magnetic interactions between such \jeff\ moments can take on a form that is different from usual Heisenberg superexchange interactions. Specifically, bond-dependent Kitaev interactions are believed to arise when these \jeff\ moments reside on a honeycomb lattice. Due to the bond-dependence of the Kitaev interaction, strong frustration exists within this model which can induce a spin-liquid ground state. \cite{Jiri2010,Khaliullin2012}

\NIO\ is a promising candidate in which Kitaev interactions might be realized. In \NIO, edge-sharing IrO$_6$ octahedra form a honeycomb net which is decorated by Ir$^{4+}$ ions with a $5d^5$ electronic configuration.  \cite{Yogesh2010} Experiments have indicated that \NIO\ is a Mott insulator with an optical gap of $\sim$350 meV. \cite{Comin} Below T = 15 K it orders antiferromagnetically in a so-called zig-zag structure.\cite{Xuerong2011} Recent resonant inelastic x-ray scattering (RIXS) results have found evidence that the large spin-orbit coupling (SOC) in \NIO\ is a dominant energy scale. \cite{Na213_dd} This would cause the Ir moments to acquire a significant orbital component, giving it a \jeff=1/2 ground state, \cite{BJKim2009} and rendering the Kitaev interaction \cite{Jiri2010,Yogesh2011} relevant.

In particular, a combination of magnetic susceptibility measurements and theoretical calculations on both \NIO\ and \LIO\ was used to claim that the spin model of this system is described as a Heisenberg-Kitaev model. \cite{Yogesh2011} However, strong second and third neighbour exchange interactions are required to account for the ordered state, the large frustration parameter, and the magnon excitations at low temperatures. \cite{Coldea2012} Indeed, recent first-principle calculations suggest a considerable electron delocalization in quasi-molecular orbits (QMO), \cite{Mazin2012,Mazin2013} which is compatible with the experimental results. At present it is thus unclear, whether a localized scenario with contribution of Kitaev interactions or alternatively a more itinerant QMO picture is more appropriate for \NIO. One of the methods to test the existence of Kitaev interactions in \NIO\ is to study its magnetic excitation spectrum. RIXS at the Ir L$_3$ edge has been successfully used to map ``magnon" dispersions in iridates.  For example, a sizable departure from a Heisenberg-like model was observed going from \SIO\ \cite{Sr2IrO4-RIXS} to \BSIO\ \cite{Sr3Ir2O7-RIXS} and was attributed to new bond-dependent magnetic interactions.

Here we present Ir L$_3$ edge RIXS results on a single crystal sample of \NIO. Our high-resolution RIXS measurements allow us to resolve new low energy excitations. In particular, at the $\Gamma$-point we observe a magnetic excitation which is centered around 35 meV. This excitation is almost dispersionless along the  {\bf Q} = [0 1 0] direction but softens along  {\bf Q} = [1 0 0]. The large magnetic intensity observed at the $\Gamma$-point is
consistent with the theoretical prediction including a bond-dependent Kitaev interaction in \NIO. \cite{Khaliullin2012} An unusually strong electron-lattice coupling is also evident in \NIO\, manifesting itself in the appearance of resonantly enhanced vibrational modes at high temperatures.

\begin{figure*}[htb]
\includegraphics[width=2\columnwidth]{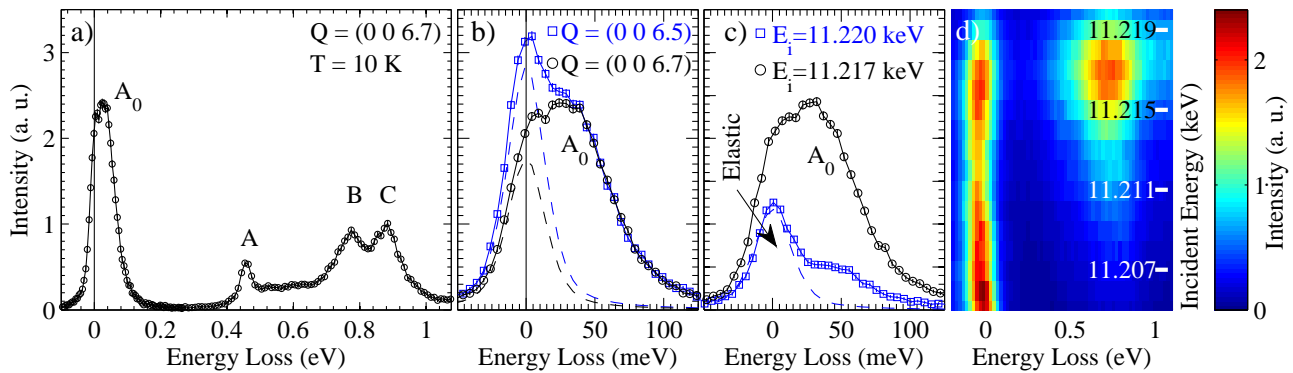}
\caption{\label{fig01}(Color online) a) Wide range RIXS spectrum for a single-crystal sample of \NIO\ at {\bf Q} = (0 0 6.7) obtained with $E_i$ = 11.217 keV. b) Detailed view of the low-lying RIXS excitations in a) taken at two different momentum transfers along the L-direction. Dashed lines are a fit to the elastic line (details provided in main text). c) and d) Incident energy dependence of the RIXS spectum. The intensity scale for (d) is shown.}
\end{figure*}

%%%%%%%%%%%%%%%%%%
%% Experimental %%
%%%%%%%%%%%%%%%%%%

The RIXS experiment was carried out at the Advanced Photon Source using the 30ID MERIX spectrometer. A spherical (2~m radius) diced Si(844) analyzer was used. The overall energy resolution (FWHM) in this configuration was $\sim$ 30 meV.  In order to minimize the elastic background intensity, most measurements were carried out in a horizontal scattering geometry near {\bf Q} = (0 0 6.75), for which the scattering angle 2$\theta$ was close to 90$^\circ$. A single crystal of Na$_2$IrO$_3$ was grown by the solid state synthesis method, as described in Ref.~\onlinecite{Yogesh2010}. The \NIO\ crystal was plate-like with a flat shiny surface, and a surface normal in the (001) direction. Throughout this paper we will use the $C2/m$ notation \cite{Coldea2012,GangCao2012} to describe the crystal structure.

%%%%%%%%%%%%%%%%%%%%%%%%%%
%% Results and Analysis %%
%%%%%%%%%%%%%%%%%%%%%%%%%%

The RIXS process at the $L_3$-edge of Ir (or any other $d$ electron
system) is a second order process consisting of two dipole
transitions ($2p \rightarrow 5d$ followed by $5d \rightarrow 2p$).
As such, it can probe excitations between the $d$-levels,
\cite{Moretti2011,Ghiringhelli,Vernay,Ghiringhelli2004,Luuk,Sr3CuIrO6,Na213_dd} collective magnetic excitations \cite{Braicovich09,Braicovich10,Sr2IrO4-RIXS,Sr3Ir2O7-RIXS} and even lattice vibrational modes.  \cite{Braicovich10,Yavas10,Steve2013,Luuk11}

In Fig. \ref{fig01} (a), a representative high-resolution RIXS spectrum of \NIO\ is plotted on a wide energy scale. This scan was
obtained at T = 10 K with {\bf Q} = (0 0 6.7) and plotted as a function of energy loss $(\omega=E_i-E_f)$. The incident energy, $E_i=11.217$ keV, was
chosen to maximize the resonant enhancement of the spectral features
of interest below 1~eV. Multiple peaks are observed between $0.4$-$1$ eV  (labeled A, B, and C). According to Ref.~\citenum{Na213_dd}, features B and C correspond to excitations between the \jeff=3/2 and 1/2 states, and feature A corresponds to an exciton formed by a particle-hole pair across the gap. \cite{Na213_dd} These features will not be discussed further in this paper. Below these peaks ($\omega < 400$ meV) we can observe the onset of the gap followed by what appears to be a large elastic line ($\omega=0$). The elastic line is plotted on an expanded energy scale in Fig. \ref{fig01} (b); an asymmetric line shape is evident with a maximum intensity at an energy loss of about 35 meV.
Moving to a different momentum transfer,  {\bf Q}=(0 0 6.5), the asymmetry remains, but since 2$\theta$ is now further away from 90$^\circ$ the intensity of the elastic line has increased. By comparing the spectra at these two {\bf Q}-points it is clear that the asymmetry is caused by new low-lying excitations, labelled A$_0$.  To emphasize this, the elastic contributions to the spectra are shown as dashed lines  using the instrumental resolution function (described later in the text).

Insights on the origin of A$_0$ can be obtained through the incident energy dependence, which reveals which unoccupied 5d states comprise the intermediate state of the RIXS process. In Fig. \ref{fig01} (c) we compare spectra taken at $E_i=11.217$ keV and $E_i=11.220$ keV. At $E_i=11.220$ keV, corresponding to exciting an Ir 2p$_{3/2}$ core electron into the unoccupied Ir 5d $e_g$ level, \cite{CuIr2S4} we notice a drastic drop in A$_0$ intensity.  In Fig. \ref{fig01} (d) we plot the incident energy dependence of the RIXS spectrum obtained with a low resolution setup (FWHM $\sim$ 150 meV). From the intensity plot we observe that A$_0$ resonates around $E_i=11.217$ keV, corresponding to the t$_{2g}$ intermediate state, just like features at higher energies. \cite{Na213_dd} In other words, the RIXS process for A$_0$, as well as those for A-C, involve an intermediate state which excites a 2p$_{3/2}$ electron into the 5d t$_{2g}$ states. The increased intensity of the elastic line below $E_i=11.211$ keV comes from the decreased absorption as the incident energy falls below the Ir L$_3$ edge.

Let us consider possible explanations for the origin of  A$_0$. We can quickly discard  excitations between \jeff\ levels since peak B represents the lowest energy transition possible. \cite{Na213_dd} In addition, the peak position of A$_0$ is an order of magnitude smaller than the optical gap, \cite{Comin} which excludes any charge related excitations. This leaves us with either magnetic or lattice excitations, \cite{Luuk11} both of which are expected to appear in this energy range. We will model the total intensity of our RIXS spectrum with 3 components:
\begin{equation}
I=I^{bg}+I^{l}+I^{m}
\end{equation}

\noindent where $I^{bg}$, $I^{l}$  and $I^{m}$ represents the elastic background, lattice vibrational modes and magnetic excitations, respectively. We use a pseudo-Voigt lineshape as the instrumental resolution function, which is a mixture of Lorentzian and Gaussian functions with equal widths and amplitudes:

\begin{equation}
%\begin{split}
 R(\omega_0,\omega) = {\Gamma^2 \over (\omega_0-\omega)^2+\Gamma^2}+e^{-(\ln 2) (\omega_0-\omega)^2/\Gamma^2} ,
%\end{split}
\end{equation}

\noindent Here the FWHM was kept at the resolution-limited value of $2\Gamma=33$ meV. For the rest of this paper we will refer to $R(\omega_0,\omega)$  as the resolution function which is centered at $\omega_0$. An example of this function is shown in Fig. \ref{fig01} (b) and (c), where the elastic line has been fit using $I^{bg}(\omega)=A^{el}\cdot R(0,\omega)$, with $A^{el}$ the amplitude of the elastic line.

\begin{figure}[htb]
\includegraphics[width=\columnwidth]{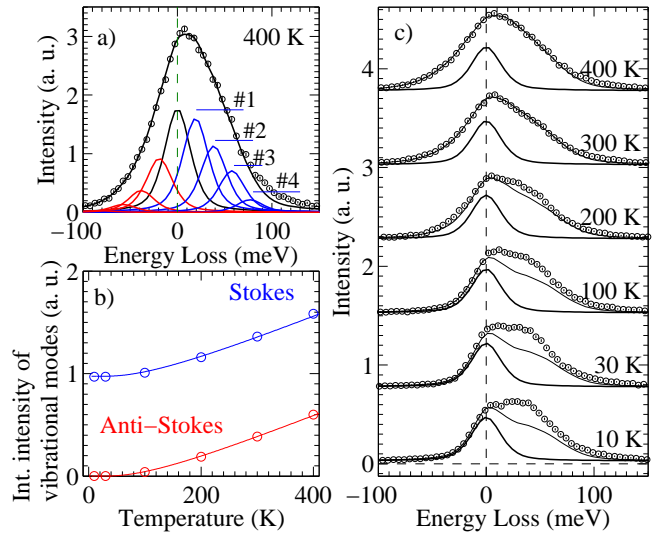}
\caption{\label{fig02}(Color online) a) Our fit function for the T = 400 K  RIXS spectrum, showing the number of resolution limited functions which represent the elastic line and the lattice vibrational modes. b) Calculated total intensity of the Stokes ($\omega > 0$ meV) and anti-Stokes ($\omega < 0$ meV) contribution to the lattice vibrations as a function of temperature. c) Temperature dependence of the RIXS spectrum taken at 10 K $\leq$ T $\leq$ 400 K for {\bf Q} = (0 0 6.7). Spectra have been shifted vertically for clarity. The thick black solid lines show the elastic component, while the thin black lines include the lattice vibrational contribution as well. }
\end{figure}

Since it is difficult to distinguish between $I^{l}$ and $I^{m}$, especially when lattice and magnetic excitations appear on the same energy scale, \cite{Braicovich10} we rely on temperature dependence to analyze the observed spectra. At temperatures well above the characteristic magnetic energy scale, lattice excitations will dominate the inelastic signal, while at lower temperatures the vibrational modes will coexist with spin excitations. In Fig. \ref{fig02} (b) we show the RIXS spectrum taken at T=400~K, well above the Curie-Weiss temperature, $\Theta_{\rm CW}$ = -116 K,  of \NIO. \cite{Yogesh2010}
At this elevated temperature only lattice vibrations are expected to contribute to the low energy RIXS signal. Note that this high temperature spectrum is much more symmetric than the T = 10 K data. The maximum intensity position seems to be closer to the elastic line due to the large spectral weight seen on the energy gain side.

In a recent RIXS study of the edge sharing cuprate compound Ca$_{2+5x}$Y$_{2-5x}$Cu$_5$O$_{10}$,\cite{Steve2013} it was found that the resonant enhancement of vibrational modes does not occur uniformly. That is, certain modes and their higher harmonics are selectively enhanced in a RIXS experiment. Motivated by this work we model the $I^{l}$ contribution with a series of vibrational modes:

\begin{equation}
I^{l}(\omega) =  \sum_{i=1}^{4} A^{l}_i \left[ (n(\omega_{i})+1)R(\omega_{i},\omega) +  n(\omega_{i})R(-\omega_{i},\omega)\right],
\end{equation}

\noindent where the $i$-th harmonic of the vibrational mode is described by a resolution limited peak centered at $\omega_{i}$. $n(\omega_i)=1/(e^{\omega_i/k_BT}-1)$ is the Bose population factor. Note that the principle of detailed balance constraints the ratio of Stokes and anti-Stokes peaks, leaving the amplitudes ($A^{l}_i$) as only adjustable parameters. Based on recent optical conductivity data \cite{Noh} we fixed our lowest mode at $\omega_1=18$ meV \cite{NaPhonon} with higher levels placed in a harmonic order \cite{Steve2013,Azcel2012,Jan2012} of the first peak ($\omega_2=36$ meV, $\omega_3=54$ meV and $\omega_4=72$ meV). In Fig. \ref{fig02} (a) we show our fit result as a solid black line using $I=I^{bg}+I^{l}$, where only $A^{el}$ and  $A^{l}_i$  were fit.

Having estimated $I=I^{bg}+I^{l}$ at T = 400 K we can proceed to calculate its expected temperature dependence for T$<$400 K by assuming that $I^{bg}$ is independent of temperature and that the intensity of the vibrational modes follows the thermal population factors: $(n(\omega_i)+1)$ and $n(\omega_i)$ for Stokes  and anti-Stokes, respectively. One can see the expected temperature dependence of the vibrational modes intensity in Fig. \ref{fig02} (b). At T = 10 K the anti-Stokes vibrational mode contribution vanishes. In Fig. \ref{fig02} (c) the calculated intensity of $I=I^{bg}+I^{l}$ is plotted as thin solid lines. The RIXS spectra shows extra intensity for T$\leq$ 200 K. This extra intensity grows as temperature decreases and leads to the appearance of A$_0$ (peaked around 35 meV).

This study of temperature dependence clearly shows that on top of the vibrational modes, there exists additional intensity that grows with decreasing temperature. The most likely origin of this intensity is magnetic for the following two reasons: 1) the temperature scale for the onset of this intensity is the same order of magnitude as $\Theta_{\rm CW}$,\cite{Yogesh2010} and 2) the ground state of this sample is magnetically ordered. \cite{Xuerong2011} We also note that these lattice vibrational modes are unusually strong, with higher-harmonics carrying appreciable spectral weight which is even comparable to the charge excitations seen in Fig. \ref{fig01} (a).  In Ir L$_3$ edge RIXS experiments electron-lattice coupling can cause lattice vibrational excitations to acquire significant spectral weight. \cite{Luuk11} This comes about due to the sudden change in the charge density around the Ir atom during the $2p \rightarrow 5d$ absorption process. The large intensity of the vibrational modes and significant spectral weight of higher-harmonics in \NIO\ is thus an indicator of strong electron-lattice coupling, \cite{Luuk11,Steve2013} which might help explain the unusual spectral broadening observed by angle resolved photoemission spectroscopy. \cite{Comin,Horsch}

\begin{figure}[htb]
\includegraphics[width=\columnwidth]{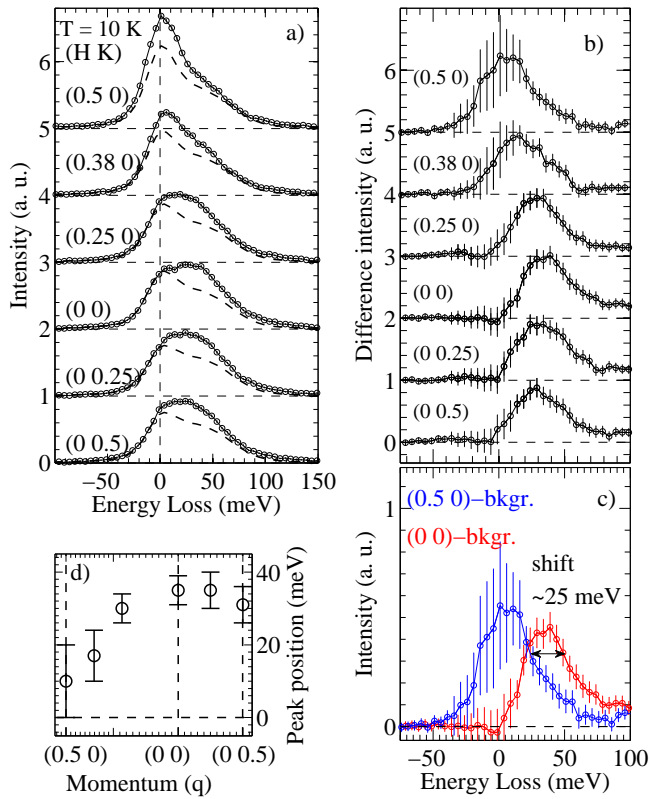}
\caption{\label{fig03}(Color online) (a) Momentum dependence of the RIXS spectra taken along the $\Gamma-X$ and $\Gamma-Y$ directions at T = 10 K. Superimposed on each spectra is the background contribution from the elastic line and the vibrational modes. (b) The magnetic signal after subtracting the background, and (c) a direct comparison between  {\bf Q} = (0 0) and {\bf Q} = (0.5 0). In (a) and (b) spectra have been shifted vertically for clarity. (d) Fit results for the peak position of the magnetic signal in (b).}
\end{figure}

We now move on to discuss the dispersion of the magnetic excitation. In Fig.~\ref{fig03} (a), the momentum dependence of  A$_0$ along both the $\Gamma-X$ and $\Gamma-Y$ directions (from (0 0) to (0.5 0)/(0 0.5))  at T = 10 K is shown, with spectra shifted vertically for clarity. Also shown in this figure as dashed lines are the contributions from ($I^{bg}+I^{l}$) in order to show the magnetic contribution clearly. We note that the intensity of the elastic line was allowed to vary between different values of {\bf Q}, which is expected in general (for a diffuse scattering intensity). The tail on the energy gain allows us to estimate the elastic intensity. This is possible because the inelastic (lattice vibrations) contribution to the energy gain side at this temperature is negligible. However, the contributions from the vibrational modes were fixed. At both {\bf Q} = (0.38 0) and {\bf Q} = (0.5 0) the spectral weight of A$_0$ shifts to significantly lower energies. In Fig. \ref{fig03}(b), the magnetic intensity has been plotted by subtracting $I^{bg}+I^{l}$ contributions from the raw spectra. The large error bars on the energy gain side reflect the uncertainty arising from the elastic background fitting. Despite this, it is clear that at the $\Gamma$-point A$_0$ forms a well defined feature which is centered around 35 meV, and that A$_0$ disperses towards lower energy along the {\bf Q} = [1 0] direction. At {\bf Q} = (0.5 0)  most of the spectral weight from the magnetic mode is only visible below 30 meV.  Fig.~\ref{fig03} (c) shows the difference spectrum obtained at  {\bf Q} = (0 0) and {\bf Q} = (0.5 0) without an offset. Although the large error bars make it difficult to extract the shift of the peak, a decrease of roughly 25 meV in energy is observed. In an attempt to capture this dispersion we fit the magnetic signal using the resolution function provided in Eq. (2). The fitted peak positions are plotted in Fig. \ref{fig03}(d), showing significant dispersion from $\Gamma$ to $X$. On the other hand, the momentum dependence along the $\Gamma-Y$ direction (from {\bf Q} = (0 0) to (0 0.5)) is much weaker (see Fig. \ref{fig03} (b) bottom spectra); no significant shift is observed. This is in stark contrast to the 25 meV dispersion observed along the $\Gamma-X$ direction.

Our observation of a high energy ($\sim$35 meV) magnetic excitation is rather surprising. In a  recent inelastic neutron scattering experiment on a powder sample of \NIO\ a magnon mode below 6 meV was identified.\cite{Coldea2012} A pure Heisenberg model with antiferromagnetic interactions and additional long range exchanges was found to be adequate in describing this result. \cite{Coldea2012} The calculated dispersion of the high energy branch at the $\Gamma$-point, however, only reaches about 5 meV (see supplemental material in Ref. \citenum{Coldea2012}), which is significantly lower than the position of A$_0$.  Given the large energy separation, it is difficult to explain both sets of data with a purely Heisenberg Hamiltonian. Recently, Chaloupka et al.\cite{Khaliullin2012} were able to explain the observed neutron data\cite{Coldea2012} by adding a Kitaev term to the Heisenberg Hamiltonian. Moreover, this additional Kitaev term would generate a high energy magnon branch. According to Ref. \citenum{Khaliullin2012}, this branch would reach 20 meV at the  $\Gamma$-point with anisotropic dispersion along the  $\Gamma-X$ and $\Gamma-Y$ directions. These predictions therefore seem quite consistent with our experimental observations in \NIO. We acknowledge that in order to extract the size of the Kitaev term, and more importantly to determine its sign, \cite{Coldea2012,Khaliullin2012} higher resolution RIXS data will be required.

%%%%%%%%%%%%%%%%%%%%%%%%%%
%% Conclusion           %%
%%%%%%%%%%%%%%%%%%%%%%%%%%

In conclusion, we have identified new low-lying excitations in \NIO\ using high-resolution Ir L$_3$ edge resonant inelastic x-ray scattering. Temperature dependence reveals two distinct modes: a dominant lattice vibrational mode at high-temperature and a magnetic excitation which appears below T = 200 K and reaches maximum intensity at T = 10 K. The vibrational excitations were fit using an optical phonon mode at $\omega_1=18$ meV, with noticable spectral weight on the next three higher-harmonics. This suggests that the electron-lattice coupling is very strong in \NIO. The magnetic mode shows peculiar momentum dependence at T = 10 K, reaching a maximum energy of $\sim$ 35 meV at the $\Gamma$-point and dispersing to lower energies along the $\Gamma-X$ direction. The observed dispersion of magnetic excitation is consistent with theoretical calculations based on a local spin model with both Heisenberg and Kitaev interactions.

%%%%%%%%%%%%%%%%%%%%%%%%%%
%% Acknowledgement      %%
%%%%%%%%%%%%%%%%%%%%%%%%%%
We would like to thank G. Khaliullin, G. Jackeli, B. J. Kim, and S. Johnston  for valuable discussions. Research at the U. of Toronto was supported by the NSERC, CFI, and OMRI. Use of the APS was supported by the U. S. DOE,
Office of Science, Office of BES, under Contract No. W-31-109-ENG-38.


\begin{thebibliography}{38}
\bibitem[{NaP()}]{NaPhonon}
\bibinfo{note}{Calculation of the eigenvector of this mode is beyond the scope
  of this paper. However, based on relatively low frequency of this mode we
  speculate that this mode involves movement of Na atoms within the honeycomb
  lattice.}
\end{thebibliography}
\end{document}